\documentclass[10pt,twocolumn,letterpaper]{article}

\usepackage{wacv}              

\usepackage{graphicx}
\usepackage{amsmath}
\usepackage{amssymb}
\usepackage{booktabs}

\usepackage{array}
\newcolumntype{L}{>{\raggedright\arraybackslash}p{4cm}} 
\newcolumntype{K}{>{\raggedright\arraybackslash}p{3cm}} 
\usepackage{multirow}
\usepackage{booktabs}
\usepackage{graphicx}
\usepackage{MnSymbol}
\usepackage{color}
\usepackage[pagebackref,breaklinks,colorlinks]{hyperref}

\usepackage[capitalize]{cleveref}
\crefname{section}{Sec.}{Secs.}
\Crefname{section}{Section}{Sections}
\Crefname{table}{Table}{Tables}
\crefname{table}{Tab.}{Tabs.}

\def\wacvPaperID{763} 
\def\confName{WACV}
\def\confYear{2025}

\begin{document}

\title{Survival Prediction in Lung Cancer through Multi-Modal Representation Learning }

\author{Aiman Farooq\\
Indian Institute of Technology Jodhpur\\
NH 62, SupuraBypass, Jodhpur\\
{\tt\small farooq.1@iitj.ac.in}
\and
Deepak Mishra\\
Indian Institute of Technology Jodhpur\\
NH 62, SupuraBypass, Jodhpur\\
{\tt\small dmishra@iitj.ac.in}
\and
Santanu Chaudhury\\
Indian Institute of Technology Delhi\\
Hauz Khas, New Delhi\\
{\tt\small schaudhury@gmail.com}
\
}
\maketitle

\begin{abstract}
Survival prediction is a crucial task associated with cancer diagnosis and treatment planning. This paper presents a novel approach to survival prediction by harnessing comprehensive information from CT and PET scans, along with associated Genomic data. Current methods rely on either a single modality or the integration of multiple modalities for prediction without adequately addressing associations across patients or modalities. We aim to develop a robust predictive model for survival outcomes by integrating multi-modal imaging data with genetic information while accounting for associations across patients and modalities. We learn representations for each modality via a self-supervised module and harness the semantic similarities across the patients to ensure the embeddings are aligned closely. However, optimizing solely for global relevance is inadequate, as many pairs sharing similar high-level semantics, such as tumor type, are inadvertently pushed apart in the embedding space. To address this issue, we use a cross-patient module (CPM) designed to harness inter-subject correspondences. The CPM module aims to bring together embeddings from patients with similar disease characteristics. Our experimental evaluation of the dataset of Non-Small Cell Lung Cancer (NSCLC) patients demonstrates the effectiveness of our approach in predicting survival outcomes, outperforming state-of-the-art methods. 
\end{abstract}

\section{Introduction}
\label{sec:intro}
Survival prediction is critical to medical care for various diseases, including cardiovascular, chronic respiratory, neurological conditions, and multiple cancers. Non-small cell lung cancer (NSCLC), which constitutes 85\% of lung cancer cases \cite{navada2006temporal}, exemplifies the importance of this task. About 25\% of lung cancer patients experience relapse post-surgery, making alternative treatment strategies essential when surgical options are not viable. Therefore, accurate survival prediction is vital for patient classification and assisting doctors in making informed treatment decisions. Recently, the prediction of survival risk has become increasingly significant. It supports treatment planning, patient staging, and monitoring, improving cancer patients' care. In the era of precision medicine, data-driven approaches and deep learning models empower healthcare professionals to make precise future outcome predictions through survival models. However, survival prediction is complex and influenced by various factors such as disease physiology, clinical data, genomics, and treatment courses. Enhanced predictions require the availability of multiple data modalities, yet datasets often lack one or more of these modalities, posing a challenge. Despite this, the available modalities still contain valuable information, making it crucial to address missing data to maximize the utility of limited data in survival prediction models.

Survival prediction for NSCLC involves CT, PET, and molecular data. Features derived from CT scans \cite{choi2021deep,xu2021survival,chen2022ct} have shown high relevance with the survival prognosis of lung cancer. While PET scans \cite{oh2021pet,lee2022predicting} and genomics data \cite{tian2017classification,zhang2020genomic} have also been linked to survival prediction studies. Traditionally, survival prediction models like random survival forest \cite{he2022artificial}, XGBoost \cite{huang2020artificial}, and Cox proportional hazards model \cite{lynch2017fuqua} rely on deriving features from imaging modalities or clinical data. With deep learning gaining popularity over the years, several Convolutional Neural Network (CNN) based models \cite{doppalapudi2021lung} have been developed for prediction, surpassing the handcrafted approaches. For instance, Huang et al. \cite{huang2022prediction} extracted features using CNN from FDDG-PET scans and constructed random forest (RF) for the prediction. For survival prediction, Lian et al. \cite{lian2022imaging} used a graph convolutional network. In particular, models such as DeepSurv \cite{katzman2018deepsurv,she2020development} and MVAESA \cite{vo2021survival} have shown promising results. However, using only a single modality doesn't allow us to capture all the biomarkers associated with the tumor. While image biomarkers provide the advantages of characterizing heterogeneous cancer in its entirety, genomic data offers insights into the underlying molecular mechanisms driving cancer progression.\par 
Several multi-modal models are also proposed to capture and leverage complementary contextual cues across diverse modalities. 
Works such as \cite{oh2023deep, hou2022radiomics, lai2020overall } proposed incorporating imaging and clinical data for improved survival prediction in lung cancer patients. MultiSurv \cite{vale2021long} was the first model to combine information from multiple domains, including clinical, genomic, and whole-slide images (WSI), for a pan-cancer dataset survival prediction. To circumvent the reliance on labels, \cite{cheerla2019deep,ding2023pathology} used similarity metrics to guide the fusion of multi-modal representations and showed improved performance.

Despite their respective advantages, most of the methods mentioned above encounter limitations due to their dependence on labeled data and lack of feature generalizability, especially in tasks involving multi-modal data. Most multi-modal methods discussed above target WSI images for prediction, with little attention to common imaging modalities such as PET and CT. Methods handling limited data labels don't consider tumor characteristics and across-patient correspondences. We propose to circumvent the label requirement and use self-supervised learning to acquire embeddings for the patient modalities, along with feature-based correspondence via contrastive learning, aiming to bolster the generalizability of visual representations. Our framework exploits inter-subject correspondence across CT, PET, and RNA-Seq data, enforced through cross-modal cluster alignment without reliance on labels, thus excelling in the specified downstream task of survival prediction. It also exhibits robust generalization and maintains predictive accuracy when RNA-Seq data is unavailable for some patients. In summary, our contribution includes :

\begin{itemize}
    \item We propose to utilize CT, PET, and molecular (RNA-Seq) information to improve survival prediction in NSCLC patients
    \item We rely not only on cross-modality relationships for a patient but also on relationships across patients based on disease similarity.
    \item We also deal with cases where certain modalities of a patient are missing and still provide adequate results.
    \item We present experimental results showing that our Multi-Modal patient embedding module and cross-patient embedding module bring significant performance gains on the NSCLC dataset.\end{itemize}

\section{Related Work}
\subsection{Survival Prediction}

Several studies have proposed models to improve survival prediction for cancer patients. Traditionally, Cox-based survival models \cite{shulman2005prognostic}, such as the linear proportional hazards model, were used across cancers; however, they need substantial feature engineering or prior medical expertise to model treatment interactions at an individual level effectively. On the other hand, nonlinear survival methods like neural networks and survival forests can naturally capture these complex interaction terms. Still, they have not yet demonstrated effectiveness as treatment recommender systems. 
DeepConvSurv\cite{zhu2016deep} proposed using CNNs for survival analysis; they used radiological scans alongside clinical data to improve predicted outcomes. Lee et al. \cite{lee2018deephit} proposed DeepHit \cite{lee2018deephit} learns the distribution of survival times without assuming an underlying stochastic process, allowing for time-varying relationships between covariates and risks. It outperformed all baselines for breast cancer survival prediction. Katzman et al. proposed a deep learning-based survival model DeepSurv\cite{katzman2018deepsurv} initially proposed for breast cancer. It outperformed state-of-the-art for nasopharyngeal \cite{liu2020deep}, lung cancer \cite{kim2019effect} as well as for glioblastoma \cite{moradmand2021role}. It extends the Cox proportional hazards model and utilizes neural networks to model complex, nonlinear relationships in survival data. \cite{huang2020deep} saw an improvement in survival prediction for cancer patients using RNA-seq data. However, all the methods use a single modality, while multi-modal data is generally present for cancer patients.

Kalakoti \cite{kalakoti2021survcnn} used multi-omics data for lung adenocarcinoma patients to stratify patients into high and low-risk categories and improve the survival prediction. While dealing with multi-modal data, \cite{wang2021gpdbn} proposed integrating both genomic data and pathological images using a bilinear network structure, which captures the interactions between the two data modalities. The network can model complex relationships and provide more accurate prognostic predictions by fusing the high-dimensional genomic features with detailed visual features from pathology images. 

 \begin{figure*}
    \centering
    \includegraphics [width=\linewidth]{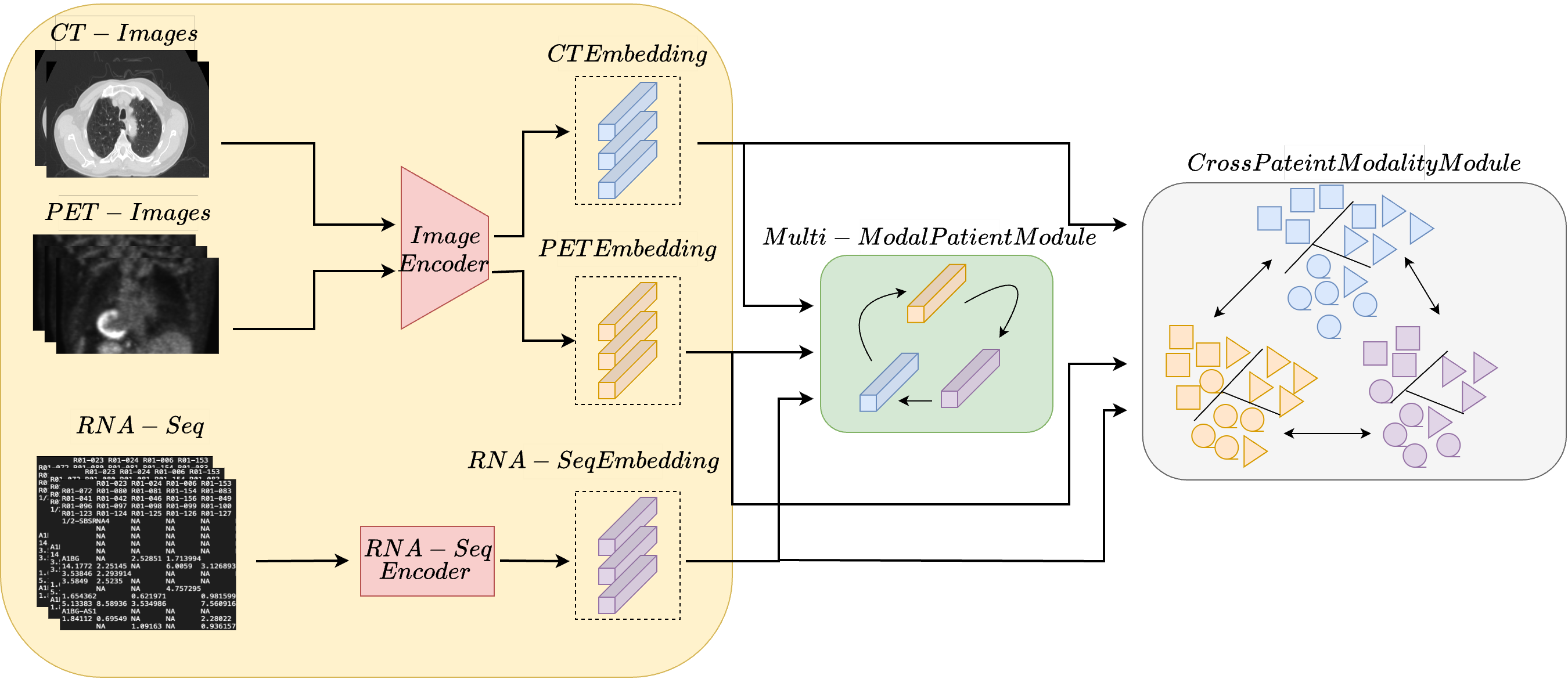}
    \caption{Overview of the multi-modal pretraining model. We show the pipeline of extracting embeddings for each modality (CT, PET, and RNA-Seq). The embeddings are refined through the MPE module, which takes care of associations between the different modalities of a patient, and the CPM module, which associates modalities across patients.}
    \label{fig:fig1}
\end{figure*}
Multisurv\cite{vale2021long} also proposed fusing learned representations from clinical, imaging, and genomic data for improved predictions across cancer datasets. Their fusion is achieved by taking the row-wise maximum, which performs better than the keyless multi-modal attention mechanism. Saeed et al. \cite{saeed2021ensemble} proposed using an ensemble model that averages out results from a CNN-based prediction model for images and a Cox proportional hazard model for clinical data. With the rise of the use of transformers for imaging tasks, \cite{saeed2022tmss} proposed encoding joint CT-PET embeddings using transformers and using the learned features for survival prediction, this model outperforms all state of the art for head and neck tumor datasets. Using both CT and PET 
 features \cite{meng2023merging} proposed a merging-diverging learning framework. The model features a merging encoder with a Hybrid Parallel Cross-Attention (HPCA) block to fuse multi-modality information through parallel convolutional layers and cross-attention transformers. The diverging decoder incorporates a Region-specific Attention Gate (RAG) block to extract region-specific prognostic information. Ding et al. \cite{ding2023pathology} proposed integrating genomic and pathology information for survival prediction in colon cancer patients. To circumvent the reliance on labels, \cite{cheerla2019deep,ding2023pathology} used similarity metrics to guide the fusion of multi-modal representations and showed improved performance.

\subsection{Survival prediction for NSCLC}
Specifically for NSCLC, much research has been done over the years to improve the survival prediction. Radiomic features \cite{van2017survival} were shown to improve the survival prediction using CT images. Sun et al.\cite{sun2018effect} studied survival prediction in NSCLC patients using Cox’s partial likelihood model. Haarburger \cite{haarburger2019image} was one of the first studies using CNNs to extract quantitative features from 3D medical images to predict lung cancer patient survival. They simplified survival analysis to median survival classification and trained the model with small batch sizes, overcoming challenges related to large batch processing and survival loss functions. \cite{nam2022histopathologic} . \cite{dao2022survival} employed a parallel transformer mechanism to capture the global context of multi-scale encoder feature maps, integrating external attention to learn the positional properties of each small patch within the dataset for segmentation, the output of which is fed to a Multimodality-based Survival Network for survival prediction in NSCLC patients. Kar et al. \cite{kar2023comparison} studied using clinical features for survival prediction using deep learning models. Li et al. \cite{li2023outcome} used features extracted from WSI using a transformer encoder to show impressive c-index values.

Several multi-modal models are also proposed to capture and leverage complementary contextual cues across diverse modalities. Oh et al. \cite{oh2023deep} proposed incorporating FDG-PET data and clinical data for improved survival prediction in lung cancer patients. Hou et al. proposed using the DeepSurv \cite{hou2022radiomics} model, which leverages radiomic features extracted from PET scans in conjunction with clinical features for the survival prediction time of NSCLC patients. Lai et al. \cite{lai2020overall} developed a Deep Neural Network (DNN) that combines gene expression and clinical data for prediction. Wang et al.\cite{wang2022novel} proposed a multi-modal deep learning model using biological knowledge that incorporated gene expression, images, and clinical data.

\begin{figure*}
    \centering
    \includegraphics[width=\linewidth]{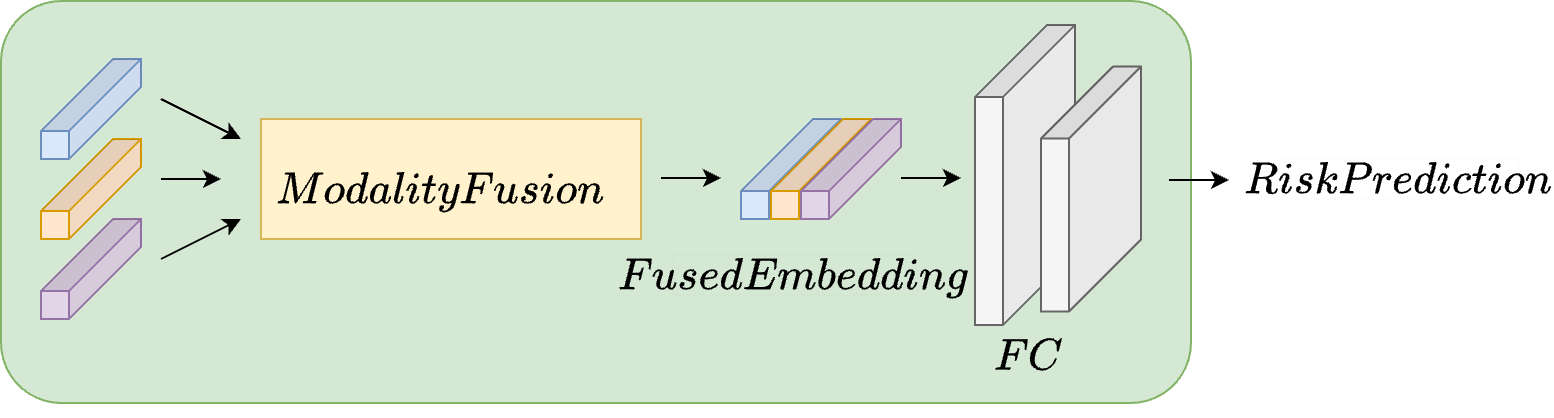}
    \caption{ Our supervised prediction model utilizes the fused feature representation and generates a conditional survival probability for each predefined follow-up time interval.}
    \label{fig:enter-label}
\end{figure*}

\section{Methodology}

Our framework consists of two main stages: a self-supervised multi-modal pretraining as shown in Fig. \ref{fig:fig1} and a supervised prediction model as shown in Fig. \ref{fig:enter-label}. The model's primary objective is to learn embeddings from imaging and genomic data to facilitate overall survival prediction. The pretraining phase captures interaction patterns among CT, PET, and genomic features, aiming to align the embeddings closely. Additionally, we leverage similarities across patients to enhance the learned embeddings. This approach allows us to extract valuable information from the dataset, even when data for specific modalities may be limited or missing.

For learning the embeddings, consider the dataset $D$ comprising of $N$ triplets of CT$(X_{tN})$, PET$(X_{pN})$ and RNA-Seq$(X_{rN})$ data of each patient. We plan on learning each embedding individually. We utilize Vision Transformer (ViT) \cite{dosovitskiy2020image} as the encoders $f_c$ and $f_p$ for CT and PET images, respectively. Additionally, for RNA-Seq embeddings, an FC net with up to 6 hidden layers $f_r$ is employed to generate the embeddings \cite{vale2021long}. Following \cite{akbari2021vatt}  each input triplet $(x_{t_i},x_{p_i},x_{r_i})$ is mapped to $(t_{i},p_{i},r_{i})$ in the latent space. Following the extraction of embeddings, our approach delves into exploiting the inherent correspondences across modalities. Our framework recognizes that each modality about a single patient inherently exhibits a natural correspondence. Building upon this, subsection \ref{ss1} presents a multi-modal patient embedding module to minimize the distances among the learned embeddings within each patient. This facilitates a closer alignment of the representations derived from CT, PET, and RNA-Seq data. Moreover, we extend our exploration to identify correspondences among the embeddings of patients with similar semantic characteristics, such as the pathological state (cancer type) in subsection \ref{ss2}. By leveraging these shared semantic features, our model can discern commonalities across patients and modalities, enriching the overall representation learning process and enhancing the predictive capabilities of our framework for survival prediction in NSCLC patients.

\begin{table*}[htbp]
\centering
\caption{Evaluation of the prediction results (C-Index) using the NSCLC-Radiogenomics Dataset}
\label{tab:table1}
\begin{tabular}{LLl}
\toprule
Model & Modality & C-index \\ \hline
\multicolumn{1}{l}{\multirow{2}{*}{XGBoost \cite{huang2020artificial}}}  & CT & 0.638 $\pm$ 0.025  \\
 & RNA-Seq & 0.610 $\pm$ 0.018 \\ 
 \midrule
\multicolumn{1}{l}{\multirow{2}{*}{RF \cite{he2022artificial}}}  & CT & 0.641 $\pm$ 0.034\\
 & RNA-Seq & 0.648 $\pm$ 0.019\\ 
  \midrule
DeepSurv \cite{katzman2018deepsurv} & CT & 0.663 $\pm$ 0.016\\
 \midrule
MCSP  \cite{nie2019multi} & CT \& PET & 0.670 $\pm$ 0.010 \\ 
 \midrule
Multimodal dropout \cite{cheerla2019deep} & CT, PET \& RNA-Seq & 0.695 $\pm$ 0.013 \\ 
 \midrule
  DeepMTS \cite{meng2022deepmts} & CT, PET  & 0.719 $\pm$ 0.040 \\ 
 \midrule
MultiSurv \cite{vale2021long} & CT, PET \& RNA-Seq& 0.713 $\pm$ 0.077\\ 
  \midrule
  XSurv \cite{meng2023merging} & CT, PET & 0.721 $\pm$ 0.063\\ 
 \midrule
   TMSS \cite{saeed2022tmss} & CT, PET \& RNA-Seq  & 0.724 $\pm$ 0.053 \\ 
 \midrule

Ours(MPE+CPM)& CT, PET \& RNA-Seq & \textbf{0.756 $\pm$ 0.020} \\ \bottomrule
\end{tabular}
\end{table*}

\subsection{Multi-Modal Patient Embedding (MPE) Module}\label{ss1}

The learned embeddings are normalized to ensure consistency across modalities. For multi-modal approaches 
 following\cite{wang2022multi,alayrac2020self}, we project the $i$-th pair, $(t_{i},p_{i},r_{i})$ into a lower dimensional embedding $(\tilde{t_{i}},\tilde{p_{i}},\tilde{r_{i}})$ and proceed to calculate the contrastive loss across each pair as:

\begin{equation}
    \ell_{i}^{ab} = -\log \frac{\exp(\text{sim}(\tilde{a}_{i} ,\tilde{b}_{i}) / \tau_1)}{\sum_{j=1}^{B} \exp(\text{sim}(\tilde{a}_{i} ,\tilde{b}_{j}) / \tau_1)},
\end{equation}

Here, $a,b \in \{t, p, r| a \neq b\}$, \(\text{sim}(\tilde{a},\tilde{b})\) calculates the cosine similarity between a pair, \( \tau_1 \) is the temperature parameter that scales the similarity values and \( B \) is the batch size. The overall loss for the MPE module is :

\begin{equation}
     L_{\text{MPE}} = \frac{1}{3N} \sum_{i=1}^{N} (\ell_i^{tp} +\ell_i^{pr} + \ell_i^{rt}).
\end{equation}

\subsection{Cross Patient Modality (CPM) Module}\label{ss2}
The MPE module allows us to align the information coming from the different modalities of a patient. However, across the patients, there may exist semantic similarity \cite{chaitanya2020contrastive,wang2022multi} based on the cancer type, which gets pushed apart due to the MPE module. We propose a Cross Patient-Modality (CPM) module to account for patient similarity. From \cite{wang2022multi}, we define $C =\{c_1,c_2,...c_K\}$, a set of $K$ trainable cross-modal prototypes and obtain the cosine similarity between each of the low dimension embedding and the cross-modal prototypes. The softmax probability ($p_{ai}$) for each modality is calculate as :
\begin{equation}
    p_{a,i}^{(k)} = \frac{\exp(sim(\tilde{a}_i,c_k) / \tau_2)}{\sum_{j=1}^K \exp(sim(\tilde{a}_i, c_{j}) / \tau_2)} ,   
\end{equation}

Here, $a \in \{t, p, r\}$ represents each of the modalities, $k$ indicates the k-th element of the vector, and $\tau_2$ is the temperature hyperparameter.
To leverage the cross-modal cross-patient similarity, we perform K-means clustering \cite{lloyd1982least} over the triplet $(\tilde{c_{i}},\tilde{p_{i}},\tilde{r_{i}})$ by individually assigning each of the modality to a cluster to obtain cluster assignments $z_{ti}$, $z_{pi}$ and $z_{ri}$ for the CT, PET and RNA-Seq embedding. The alignment across the clusters is done by the cross entropy loss as follows: 
\begin{equation}
    \ell(\tilde{a}_{i}, z_{b,i}) = \sum_{k=1}^{K} z_{b,i}^{(k)} \log p_{{a,i}}^{(k)} , 
\end{equation}

Here $a,b \in \{t,p,r| a \neq b\}$. Thus, the loss across the three modalities for the cross-modality patient module becomes :
\begin{equation}
    L_{\text{CPM}} = \frac{1}{3N} \sum_{i=1}^{N} (\ell(\tilde{t}_i, z_{pi}) +\ell(\tilde{p}_i, z_{ri}) + \ell(\tilde{r}_i, z_{ti}).
    \end{equation}
The overall loss for the pretraining is calculated as the weighted sum of the above two losses:
\begin{equation}
\label{overll}
    L_T = \alpha_1 \cdot L_{MPE} +  \alpha_2 \cdot L_{CPM}.
\end{equation}

Here, $\alpha_1$ and $\alpha_2$ are hyperparameters that determine the relative importance of each loss term in equation \ref{overll}.

\subsection{Modality Fusion and Prediction}

Following the pretraining, the embeddings can be utilized for several downstream tasks. Here, specifically, we are targeting survival prediction. We obtain a multi-modal fused representation by concatenating the embeddings obtained from the previous task, as shown in Fig. \ref{fig:enter-label}. This fused output serves as input to the FC layer for the final prediction using the Cox Proportional Hazard \cite{cox1972regression} model
\begin{equation}
  L_{\text{risk}} = - \sum_{i=1}^{N} \left[ \delta_i \left( \beta^T f_{i} - \log \left( \sum_{j \in R_i} \exp(\beta^T x_j) \right) \right) \right]
\end{equation}

Where $f_{i}$ is the fused embedding for individual $i$, $\delta_i$ is an indicator variable taking the value  $1$ if individual $i$ experienced an event, and $0$ if it was censored, $R_i$ is the set of individuals who are at risk at the same time as individual $i$ and $\beta$ is the vector of coefficients learned by the model.

In most lung cancer datasets, while the imaging modalities may be easily accessible, the genomic data may not be available due to cost constraints. Our model allows us to effectively leverage the relationships between the available modalities to impute missing values. To mimic said behavior, we replace the RNA-Seq modality with an average of the other two modalities. As a result, the reliance on complete genomic information for survival prediction is alleviated.
\section{Experiments} 
\subsection{Dataset } In this work, we use a public radiogenomics dataset of NSCLC Radiogenomics \cite{bakr2017nsclc,bakr2018radiogenomic} available in The Cancer Imaging Archive (TCIA). The imaging and clinical data are already anonymized, along with all the ethical clearances. The data includes CT images with tumor segmentation on the CT image, tumor characteristics, PET scans, and RNA-Sequence data. It contains data from about 211 patients, out of which genomic data is available for a subset of 130 subjects. For the pre-processing, CT \& PET scan slices were resampled with a slice thickness of 1 mm$^3$ and a set number of 256 slices. Standard normalization operation was performed for both modalities. 
\subsection{Implementation details \& Evaluation metric}
For the NSCLC dataset, we propose a 4-fold cross-validation setup for training and evaluation purposes. In all runs, we computed the average performance on the test set based on the results obtained from the best-performing models. We use the cross-validated concordance index (C-Index) to assess the predictive performance. The  modified C-Index \cite{tang2023pre} evaluates the ability of the model to rank accurately predicted patient risk scores in relation to overall survival as : 

\begin{equation}
C\text{-}index(D) = \frac{\sum_{i,j \in D} 1_{T_i < T_j} \cdot 1_{R_i < R_j} \cdot \delta_i}{\sum_{i,j \in D} 1_{T_i < T_j} \cdot \delta_i}
\end{equation}

Here, $T_i$ and $T_j$ represent the survival time of individual $i$ and $j$, $R_i$ and $R_j$ represent the predicted risk score of individual $i$ and $j$, $\delta_i$ is the censoring indicator for the individual and $1_{T_i < T_j}$ is 1 if $T_i < T_j$, else 0. Kaplan-Meier survival curves analysis is done to evaluate the differences in event-free survival distribution. The number of epochs is kept at 100, and the batch size is 4. The learning rate is $0.0001$ for pretraining and $0.00005$ for the prediction task. For the pretraining setup, the dimensionality is set to 128, $\tau_1$ = 0.1, $\tau_2$ = 0.2, and 
 $k$= 500. We have trained our model using NVIDIA A100 Tensor Core and RTX 3090 GPU using the  Pytorch framework.

For comparison purposes we consider state-of-the-art survival models, including TMSS\cite{saeed2022tmss} , XSurv\cite{meng2023merging}, DeepSurv \cite{katzman2018deepsurv} and MCSP \cite{nie2019multi} as baselines. For TMSS\cite{saeed2022tmss}, we replace the EHR embeddings with RNA-seq data to study the impact of all three modalities. Furthermore, due to ML-based models' success in the survival prediction task, we include RF\cite{he2022artificial} and XGBoost \cite{huang2020artificial} also in the baselines. To compare against the existing multi-modal methodologies, we include MultiSurv \cite{vale2021long} and Multi-modal dropout \cite{cheerla2019deep}. We have used the same split, hyperparameters, and pre-processing strategies to compare all models fairly.

\begin{figure}
    \centering
    \includegraphics[width=\linewidth]{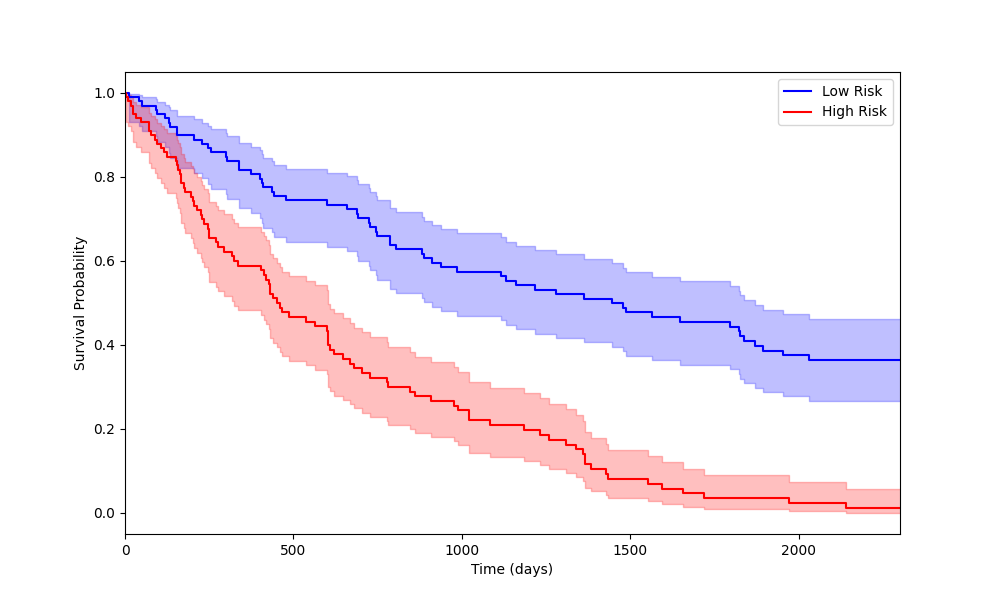}
    \caption{ Kaplan-Meier curves of identified low-risk and high-risk patients for survival prediction.}
    \label{fig:kap}
\end{figure}
\section{Results and Discussion}
Our method demonstrates improved survival prediction on the NSCLC Radiogenomics dataset as shown in Tab.\ref{tab:table1}. Our self-supervised data fusion methodology outperforms all state-of-the-art methods, including all uni-modal and multi-modal baselines. Our best-performing model is achieved by a combined multi-modal approach using CT, PET, and RNA-Seq data; the details of other combinations are presented in the ablation study subsection. We also study the impact of missing modality, particularly RNA-Seq, on the overall results. The performance sees a slight dip by replacing the missing modality with the average of the other two modalities. Specifically, even when RNA-Seq data for 10\% of patients is unavailable, our model attains a C-index value of 0.725, an improvement over existing multi-modal approaches. This shows that the model can generate meaningful embedding even with limited RNA-Seq data available. The Kaplan-Meier analysis is provided in Fig. \ref{fig:kap}

There are some limitations to our study. Mainly due to the limited nature of datasets publicly available with all three modalities available, we are unable to test our approach across multiple cancer variants, which hinders the model's generalizability. Also, a significant drop in the C-index is observed when we replace around 40\% of RNA-seq. This suggests that for larger amounts of missing data, our model is not able to achieve a good embedding.

\begin{table}[htbp]
    \centering
    \caption{Ablation study on the impact of individual modalities}
    \label{tab:table2}
    \begin{tabular}{@{}cccc@{}}
        \toprule
        \multicolumn{3}{l}{Data Modalities} & \multirow{2}{*}{C-index} \\ \cmidrule(r){1-3}
        CT & PET & RNA-Seq &  \\ \hline
        $\checkmark$ & $\checkmark$ & $\checkmark$ & 0.756 $\pm$ 0.016   \\
         \midrule
        $\checkmark$ & $\checkmark$ & - & 0.673 $\pm$ 0.021\\
         \midrule
        $\checkmark$ &- &$\checkmark$ &  0.728 $\pm$ 0.018\\
         \midrule
        -& $\checkmark$ & $\checkmark$ & 0.613 $\pm$ 0.025\\
        \bottomrule
    \end{tabular}
\end{table}
\subsection{Ablation Study}
To evaluate the impact of individual modalities during pretraining, we conducted ablation studies where we trained the model using varying combinations of the modalities. Tab. \ref{tab:table2} shows that the combination of CT, PET, and RNA-Seq outperforms all the rest. In a two-pair scenario, where the third modality is omitted altogether, the combination of CT and RNA-Seq is the best performer. To study the effect of the missing modality, we train models with varying percentages of missing RNA-Seq values in Tab. \ref{tab:table3}; we observe that even with increasing percentages of missing modality, our model continues to generate embeddings strong enough to predict accurate survival status. We also tried imputing the missing modality with zero value (dropout) and a value predicted from the available two modalities. The model yields a C-index of 0.692 with the predicted value, while zero-embedding gives a 0.673 score for 20\% missing values.

\begin{table}[htbp]
    \centering
    \caption{Ablation study on the impact of missing RNA-Seq modality}
    \label{tab:table3}
    \begin{tabular}{@{}cc@{}}
        \toprule
        \begin{tabular}[c]{@{}c@{}}\% of RNA-Seq\\ replaced\end{tabular} & C-index \\
        \hline
        10  & 0.725 $\pm$ 0.013\\
         \midrule
        20 & 0.715 $\pm$ 0.021\\
         \midrule
        30 &  0.697 $\pm$ 0.027\\
         \midrule
        40 & 0.681 $\pm$ 0.022\\
        \bottomrule
    \end{tabular}
\end{table}

\section{Conclusion}
This study presents a novel approach for predicting survival in non-small cell lung cancer (NSCLC) patients by leveraging comprehensive information from CT and PET scans and genomic data. Using the Cross Patient Modality module, our proposed model is designed to capture associations between multiple modalities for individual patients while aligning embeddings closely based on semantic similarities across patients. This work presents the first instance of utilizing multi-modal representation learning to predict overall survival in lung cancer patients. Our experimental evaluation of the NSCLC dataset patients demonstrated a remarkable 9\% increase in predictive performance compared to uni-modal approaches. Furthermore, our method exhibited a significant 3\% improvement compared to existing multi-modal approaches. Our model achieved competitive C-index scores even in scenarios with RNA-Seq modality missing, highlighting its robustness. This capability is particularly crucial when genomic data for specific subjects is unavailable. Overall, our proposed approach holds promise for enhancing patient outcomes and guiding treatment planning in the battle against lung cancer, the leading cause of cancer-related deaths worldwide.\\

{\small
\bibliographystyle{ieee_fullname}
\bibliography{ref}
}

\end{document}